# Getting the picture - Subtractive color mixing with an inkjet printer

David Marasco, Foothill College

When teaching colors, physics teachers ask their students to ignore what is taught in art classes and embrace a primary color system of red, green, and blue.[1-2] This scheme is additive and all colors can be generated by combining light of these three colors with varying intensities. There have been a number of demonstrations and tools to help students understand additive color mixing, using LEDs, color monitors, or similar technologies..[1,3-5] However, the subtractive primaries of cyan, magenta, and yellow have received far less attention, if students have previous knowledge it is from printer ink cartridges. This gap may exist because as pigments it can be harder to vary their intensities in real time.[1,2,4] Fortunately, our colleagues in the art department are well-versed in subtractive color mixing, and have taught this topic for many years through the lens of color printing. We steal a page from their book and describe how to create a color image using cyan, magenta, yellow, and black layers (CMYK). Black is used as an additional layer due to mixing issues concerning the imperfections of the CMY pigments.[6] With imaging software, an inkjet printer, and some transparencies, physics instructors are just a few keystrokes away from being able to demonstrate subtractive color mixing.

As it is widely available and free, GIMP[7] is used as the image manipulation software. GIMP assumes the user is designing for the screen, and hence by default works in the additive RGB color space. After opening the image of interest in GIMP, the next step is to separate the image into the four layers corresponding to CMYK. Click **Colors>Components>Decompose** to open the appropriate dialog box and change the color space from RGB to CMYK, and click OK. This creates four layers, each corresponding to a channel with the desired color information. Note that while these layers are labelled with the different colors, as they give intensity information they are expressed in black and white. While GIMP is assuming additive colors, we are forcing a subtractive color scheme, therefore we need to invert the values in these layers, using **Colors>Invert** for each layer (for those familiar with film photography, the layers should resemble negatives).

The resulting layers give information about how the printer should deposit ink. In white regions it will deposit no ink, and in black regions it will lay down more ink. As the eventual product will consist of actual layers of cyan, magenta, yellow, and black, the document must be converted back into color by selecting **Image>Mode>RGB**. In the cyan layer the foreground color should be set to (00FFFF, or 0% red, 100% green, 100% blue) and the background color should be white (FFFFFF). Once that is done choose **Colors>Map>Gradient Map** to convert the layer to cyan. Do the same for the magenta and yellow layers using (FF00FF) and (FFFF00) as the foreground colors. The black layer is fine, as it is already in black and white.

To construct the resulting color image, print each layer individually on a transparency. While it is best practice to put aligning marks in each layer to aid in lining up the four transparencies, I have found this unnecessary for this exercise. The transparencies are stacked on top of a piece of white paper from lightest ink to darkest (YMCK) to get a color photograph. If your school still has classic overhead projectors available they can also be used.

A sample is shown in figure 1. The blue shirt in the original photo has shifted to a purple hue. As a print of the original picture results in the proper blue, the color shift is most likely due to issues in the initial decomposition into the CMYK layers. While exact color matching when translating from screen to print is often problematic, the point of this demonstration is to show subtractive color mixing rather than create a true-color image. The tartan scarf is an homage to the very first color photo, the tartan ribbon, created by Thomas Sutton at the behest of James Clerk Maxwell.[8] Maxwell was a pioneer in our understanding of red, green, and blue as primary colors.

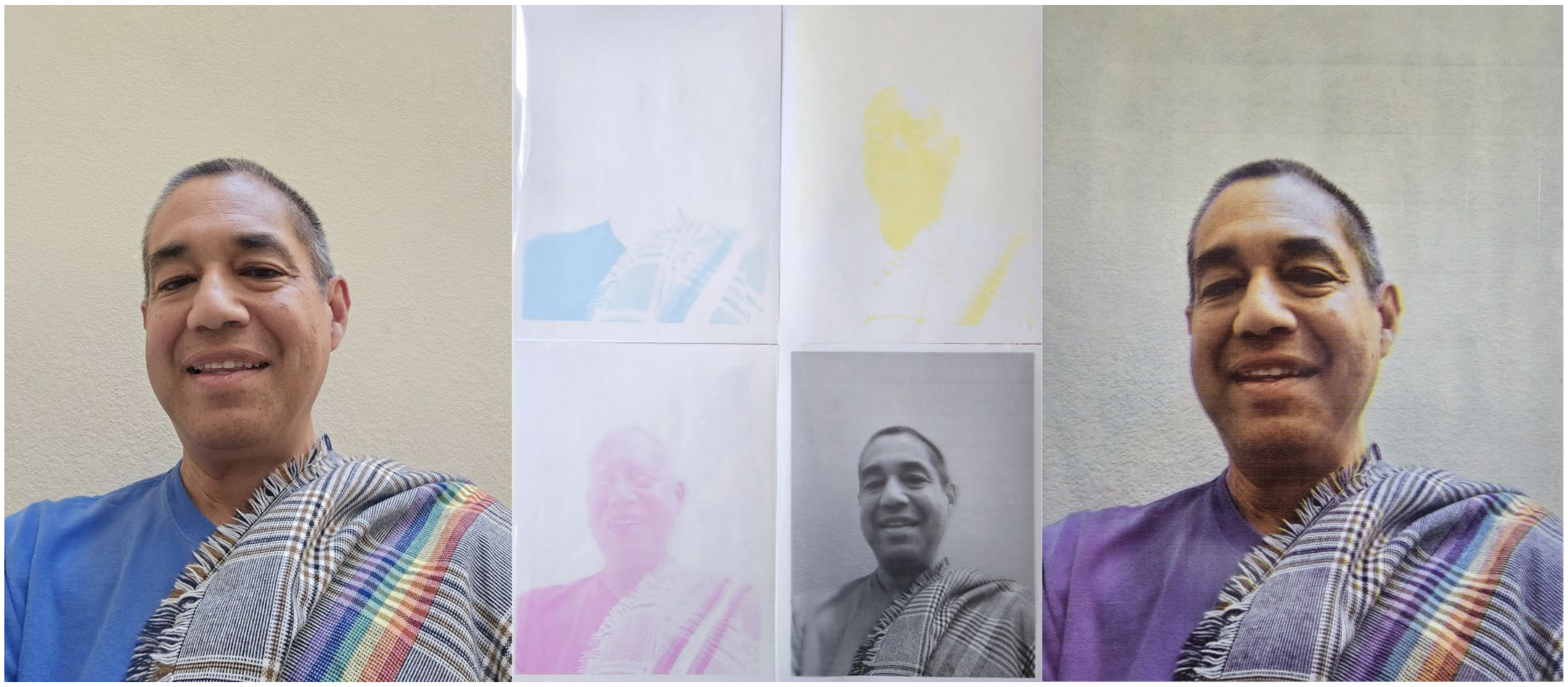

*Figure 1. The original photo is on the left, the CMYK transparencies are in the center, and the stacked image is on the right.*

Acknowledgments
The author would like to thank Bree Barnett Dreyfuss for the gift of the colorful scarf.

References:

1. Leonard Parsons; As easy as R, G, B. *Phys. Teach.* 1 September 1998; 36 (6): 347–348. https://doi.org/10.1119/1.880105

2. Lawrence D. Woolf; Confusing color concepts clarified. *Phys. Teach.* 1 April 1999; 37 (4): 204–206. https://doi.org/10.1119/1.880230

3. Gorazd Planinšič; Color Mixer for Every Student. *Phys. Teach.* 1 March 2004; 42 (3): 138–142. https://doi.org/10.1119/1.1664377

4. Frances Ruiz, Michael J. Ruiz; Color Addition and Subtraction Apps. *Phys. Teach.* 1 October 2015; 53 (7): 423–427. https://doi.org/10.1119/1.4931012

5. Rucheng Dai, Zhongping Wang, Zengming Zhang, Lazhen Sun; A Color-Mixing Setup for a Physics Teaching Laboratory. *Phys. Teach.* 1 November 2022; 60 (8): 677–680. https://doi.org/10.1119/5.0039994

6. Marvin Hoshino; A response to ’’Dilemma of the primary colors’’. *Am. J. Phys.* 1 July 1979; 47 (7): 573. https://doi.org/10.1119/1.11756

7. https://www.gimp.org/.

8. Susanne Klein and Paul Elter; The Tartan Ribbon or Further Experiments of Maxwell’s Disappointment/Sutton’s Accident. *Heritage* 2023 6(2): 968-978 https://doi.org/10.3390/heritage6020054